\pdfoutput=1

\documentclass[11pt]{article}

\usepackage[]{acl}

\usepackage{times}
\usepackage{latexsym}

\usepackage[T1]{fontenc}

\usepackage[utf8]{inputenc}

\usepackage{microtype}
\usepackage{graphicx}

\usepackage{inconsolata}
\usepackage{algorithm}
\usepackage{algorithmic}
\usepackage{amssymb}
\usepackage{newlfont}
\usepackage{amsmath}
\usepackage{booktabs}
\usepackage{adjustbox}
\usepackage{diagbox}

\newcommand{\subheadi}[1]{\vspace{0.3\baselineskip}\noindent\textit{#1}}

\title{ \vspace*{-0.5in}
{{\small \hfill ACL'24}\\
\vspace*{.25in}}
Is Table Retrieval a Solved Problem?\\ Exploring Join-Aware Multi-Table Retrieval}

\author{Peter Baile Chen \\
  MIT \\
  \texttt{peterbc@mit.edu} \\\And
  Yi Zhang\thanks{Work was done before joining AWS.}\\
  AWS AI Labs \\
  \texttt{imyi@amazon.com} \\ \And
  Dan Roth \\
  University of Pennslyvania\\
  \texttt{danroth@seas.upenn.edu}
  }

\begin{document}
\maketitle
\begin{abstract}



Retrieving relevant tables containing the necessary information to accurately answer a given question over tables is critical to open-domain question-answering (QA) systems.
Previous methods assume the answer to such a question can be found either in a single table or multiple tables identified through question decomposition or rewriting. 
However, neither of these approaches is sufficient, as many questions require retrieving multiple tables and joining them through a \textit{join plan} that cannot be discerned from the user query itself. 
If the join plan is not considered in the retrieval stage, the subsequent steps of reasoning and answering based on those retrieved tables are likely to be incorrect.
To address this problem, we introduce a method that uncovers useful join relations for any query and database during table retrieval. We use a novel re-ranking method formulated as a mixed-integer program that considers not only table-query relevance but also table-table relevance that requires inferring join relationships.  Our method outperforms the state-of-the-art approaches for table retrieval by up to 9.3\% in F1 score and for end-to-end QA by up to 5.4\% in accuracy.\footnote{Data and code are available at \url{https://peterbaile.github.io/jar/}.}

\end{abstract}

\section{Introduction}
\label{sec:intro}
\begin{figure*}
    \centering
    \includegraphics[width =0.8\textwidth, trim=0in 0.3in 0in 0.1in,clip]{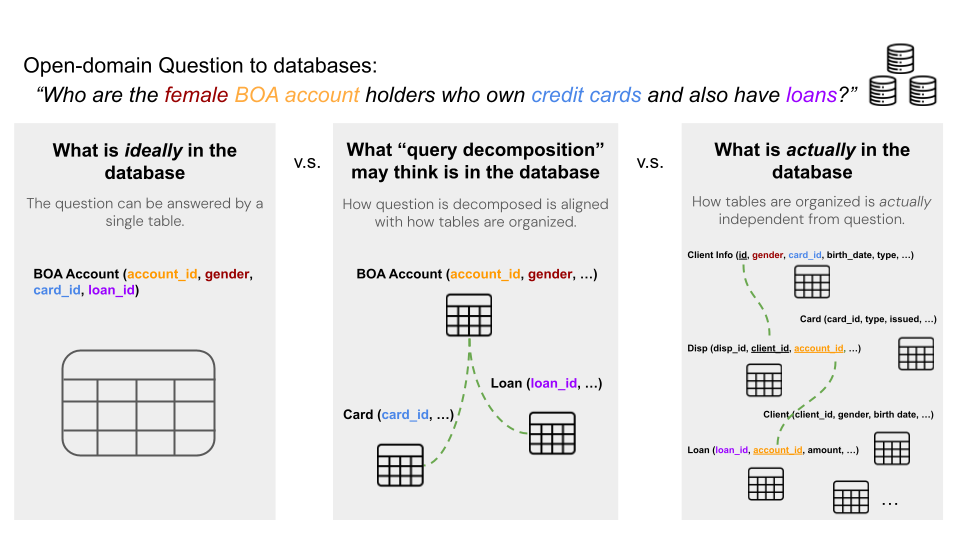}
    \caption{
    Previous work makes simplifying assumptions for table retrieval. Left block assumes there exists a single table that is sufficient to answer the question. Middle block assumes necessary joins can be recovered from query decomposition.
    However, in practice, the question may be more complicated and the join plan may not be discerned from the user query itself (right block). Therefore it is important to identify the join relationship while conducting table retrieval.}
    \label{fig:intro_exp}
\end{figure*}

Structural tables are an important source of knowledge for many real-world applications, such as open domain question answering~\cite{opentableqa} and open fact-checking~\cite{schlichtkrull-etal-2021-joint}. A standard paradigm to leverage structural tables as a knowledge source follows first retrieving relevant tables, then conducting standard question answering or fact verification over those top-ranked tables. 
This setup has been recently adopted by Retrieval Augmented Generation (RAG)~\cite{rag,pan2022end, gao2023retrieval}, trying to address the issue of knowledge update~\cite{yu2022survey} and hallucination~\cite{maynez-etal-2020-faithfulness,zhou-etal-2021-detecting} of Large Language Models (LLMs).
Once the tables required to answer the user prompt are identified, they will then be fed to LLMs with either zero-shot prompts or in-context learning examples~\cite{gpt3} to generate the final answer. 
Since the reasoning and answer generation is conditioned on the retrieved tables, table retrieval is a critical problem.
Recent works~\cite{opentableqa,table-text} have proposed to learn a model that can capture the similarity between table contents and the given query using either a text or table-specific model design. 
However, these works simplify the table retrieval problem in real-world open-domain settings. The reasons are as follows.

First, to answer a question or verify a fact relevant to structural tables, we may need \textit{multiple} tables.
For example, given a database and a query
\textit{"Who are the female BOA account holders who own credit cards and also have loans?"}, ideally, we may want to find an account table with a schema including columns such as \texttt{account\_id}, \texttt{gender}, \texttt{card\_id}, \texttt{loan\_id}, as shown in the first block in Figure~\ref{fig:intro_exp}, which we can directly use to answer the query by writing a SQL query. However, in real-world databases, it is not very likely that all the information is stored in a single table. In our running example, since a client can have multiple credit cards and loans, to avoid repeating information which leads to a very large table, the single table is typically split into three separate joinable tables, i.e., an \texttt{Account} table, a \texttt{Card} table, and a \texttt{Loan} table. All these tables need to be retrieved to answer the query correctly. Therefore, table retrieval should not only target retrieving a single relevant table but always \textit{multiple} tables.

Secondly, retrieving multiple tables requires understanding the \textit{relationship} among candidate tables, which is independent of how the query is framed.
A straightforward approach to solve multi-table retrieval can be training a model that can decompose queries~\cite{schick2023toolformer}, so that each query is aligned to a table. 
In our running example, the query mentions \texttt{account}, \texttt{credit cards}, and \texttt{loans}, which may indicate the set of required tables.
Although the decomposition aligns with how the query is framed, it might not align with how relevant data are organized and stored.
For instance, relevant tables might have the following structures:

\textit{Client Info} (\texttt{id}, \texttt{gender}, \texttt{card}\_\texttt{id}, ...)

\textit{Disp} (\texttt{disp\_id}, \texttt{client\_id}, \texttt{account\_id} ...)

\textit{Loan} (\texttt{loan}\_\texttt{id}, \texttt{account}\_\texttt{id}, \texttt{amount}...)

In this case, we may need to join all of them, as shown in the third block in Figure~\ref{fig:intro_exp}, to answer the given query.
However, we may also find another set of tables, for example:

\textit{Card} (\texttt{card\_id}, \texttt{type}, \texttt{issued}, ...)

\textit{Client} (\texttt{client\_id}, \texttt{gender}, \texttt{birth date}, ...)

\textit{Loan} (\texttt{loan}\_\texttt{id}, \texttt{client}\_\texttt{id}, \texttt{amount}, ...)

Although this set of tables seems to be highly relevant, it is not the right set of tables to return because they cannot be joined. In particular, the relationship between the cards listed in the \texttt{Card} table and the clients in the \texttt{Client} table is missing, and therefore they can not produce the answer. Importantly, only if it understands the relationship among tables can the model determine which set of tables to return.

Last but not least, the relationship among tables needs to be \textit{inferred}. Previous works in Text-to-SQLs~\cite{yu-etal-2018-spider,li2023can, lee-2021-kaggle-dbqa} have considered table joins when generating SQL expressions. However, they all assumed that the key-foreign-key constraints (join relationships) had already been provided with the database. In reality, this is not always possible, in particular when we are searching from data lakes~\cite{josie,datalake,juneau,cong2022warpgate} or simply a big dump of tables, which introduces additional challenges during the retrieval. 

To sum up, the problem of table retrieval was simplified in previous works, whereas it can be more challenging in a real-world open-domain setup.
To solve the problems we have discussed, in this paper, we formulate the multi-table retrieval problem as a re-ranking problem over a ranked list of tables produced by a standard table retrieval method, and we developed an algorithm to select the best set of tables during test time by considering both relevance and join relationship together. We conduct extensive experiments on Spider~\cite{yu-etal-2018-spider} and Bird~\cite{li2023llm} datasets, demonstrating that our re-ranking mechanism outperforms baselines in both retrieval and end-to-end performances.



\section{Problem Description}


A standard table retrieval problem for an open-domain question-answering task over tables can be defined as follows. Given a query $Q$, a table corpus $C = \{T_i\}^M_{i=1}$, retrieve a table $T_i$ from $C$, such that the answer of $Q$ resides in $T_i$.
This paper extends the problem definition to consider multiple tables that require table joins. 

Specifically, given a query $Q$, a table corpus $C$, rather than returning a single table $T_i\in C$, our problem is to return a ranked list of tables with join operators denoted as $E(Q)$,
such that the answer of $Q$ resides in the tables in $E(Q)$. Here $E(Q) =\{T_{Q,1} \Join_{c_1} T_{Q,2} , ...\}$
, where $T_{Q,1}, T_{Q, 2}, ... \in C$, and $c_1, ...$ represent the join conditions.
For example, if table \texttt{Client Info}'s column \texttt{id} should join table \texttt{Disp} on its column \texttt{client\_id}, then $c_i = (id, client\_id)$. 

\section{Join-aware Multi-Table Retrieval}
\label{sec:re-ranking}
To address the problem described in the section above, i.e., identifying the best table expression over a subset of tables from the given databases, we need to consider two aspects. One is \textit{table-query relevance}, which estimates how likely a candidate table contains the information that can be used to answer the query. The other one is \textit{table-table relevance}, which is to identify the tables that contain complementary information belonging to the same objects. In this case, \textit{table-table relevance} is equivalent to how likely tables can be joined.

We note that these two components should be considered simultaneously to answer the query because only tables that are relevant to the query and compatible with each other can provide sufficient information to answer the query.
Therefore, we propose to solve the problem by adjusting the ranking produced by \textit{table-query relevance} based on relationships inferred from \textit{table-table relevance} to find the best retrieval result. 

In the following sections, we will focus on this re-ranking problem, developing a computational framework that can consider and infer the join relationship while generating the new table ranking to help answer the query.

\subsection{Query-Table Relevance}
\label{sec:qt-rel}
A standard way to compute query-table relevance is to compute a similarity score through a bi-encoder model architecture~\cite{chen2020table,huang2022mixed,opentableqa}.
Such a similarity score can capture coarse-grained relevance. However, real queries can be complicated (e.g., multi-hop questions). The relevant information to answer the query may be distributed across multiple tables, and each table that should be used to answer the query may be only mapped to a part of the query. In our running example, a \texttt{Card} table may only tell us \textit{"who own credit cards"} but not \textit{"who have loans"}.

Therefore, we argue that fine-grained query-table relevance scores are also necessary so that we can find relevant tables (and their columns) for different information that is required to answer the query. To achieve this, we compute the score by first decomposing the query and then computing the semantic similarity between each sub-query and columns from candidate tables.
Specifically, for a query $Q$, we want to decompose it into sub-queries that can potentially be mapped to tables and columns. To achieve this goal, rather than simply using keywords, we use \textit{concept} and its \textit{attribute} mentioned in $Q$ as a pair to represent a sub-query.
For instance, the query \textit{``What is the id of the trip that started from the station with the highest dock count?''} requires information relevant to two main concepts: trip and station. More specifically, the query needs information about the attribute ``id'' from the trip concept and the attribute ``dock count'' from the station concept. Therefore, the query should be decomposed into $\langle$trip, id$\rangle$ and $\langle$station, dock count$\rangle$, which may better indicate that we need to look for a table about trip with id information, and a table about station with dock information respectively. 

To perform the decomposition, we use an LLM (GPT-3.5 Turbo for our experiment) directly by feeding the prompt stated in \ref{app:prompt-decomp} to it.\footnote{We used 5 ICL examples.}
Then, for each sub-query $q \in Q$, the similarity $r_{qik}$ between $q$ and column $c_k$ of table $T_i$ is calculated as the semantic similarity between $q$ and column $c_k$ using a standard bi-encoder model~\cite{contriever}.

\subsection{Table Re-ranking with Table-Table Relevance}
\label{sec:mip}
As discussed above, we prefer tables with high \textit{query-table relevance} and \textit{table-table relevance}.

For query-table relevance, we consider both \textit{coarse-grained} and \textit{fine-grained} relevance introduced in Section~\ref{sec:qt-rel}, so that we can find the tables that are most relevant to all the information that is required to answer the query. To ensure the coverage of different information, we further complement the relevance score with a coverage measurement, which will be described in detail in Section~\ref{sec:coverage}. 

For \textit{table-table relevance}, we consider the compatibility among the returned tables, i.e., whether selected tables can join with each other. 
The compatibility (joinability) helps us ensure that the information from the selected tables is compatible --- the information is about the same group of entities or objects, so that we can use \textit{all of them} to answer the query correctly. Therefore, during the process of selecting tables that can maximize the compatibility scores, we further encourage the selected tables to be chained through join, which will be discussed in Section~\ref{sec:connect}.


All in all, we formulate the re-ranking problem as an optimization problem, maximizing the sum of the aforementioned components as a Mixed-integer Linear Program (MIP). To incorporate the coarse-grained relevance scores, we define the binary decision variable $b_i$ that denotes whether table $T_i\in C$ is chosen to be returned and parameter $r_i$ that denotes the coarse-grained relevance score between the input query and table $T_i$. The maximization term, the total coarse-grained relevance of tables selected, is $\sum_i r_i b_i$. Similarly, to consider the fine-grained relevance scores, we introduce the binary decision variable $d_{qik}$ that denotes whether or not sub-query $q$ is covered by column $c_k$ of table $T_i$ and parameter $r_{qik}$ that denotes the fine-grained relevance score between sub-query $q$ and column $c_k$ of table $T_i$. The maximization term, the total fine-grained relevance of each sub-query, is $\sum_{q, i, k} r_{qik} d_{qik}$. To take into account the compatibility scores, we define the binary decision variable $c_{ij}^{kl}$ that describes whether column $c_k$ of table $T_i$ is selected to join with column $c_l$ of table $T_j$ and parameter $\omega_{ij}^{kl}$ that denotes the compatibility between column $c_k$ of $T_i$ and column $c_l$ of table $T_j$. The maximization term, the total pair-wise compatibility of columns selected, is $\sum_{i, j, k, l} \omega_{ij}^{kl} c_{ij}^{kl}$.


\subheadi{Objective.}
Using the above variables, parameters, and three maximization terms, we define our objective function to return the best $K$ tables in the following equation
\begin{equation*}
\label{eqn:obj}
\arg \max \sum_i r_i b_i + \sum_{q, i, k} r_{qik} d_{qik} + \sum_{i, j, k, l} \omega_{ij}^{kl} c_{ij}^{kl}
\end{equation*}
The objective function is subject to the following six constraints: 

\subheadi{Constraint 1.} The integral constraint ensures the decision variables are binary, as defined above.
\begin{equation}
b_i, c_{ij}^{kl}, d_{qik} \in \{0, 1\} 
\end{equation}

\subheadi{Constraint 2.} This constraint sets the maximum number of tables and join relationships that can be selected according to the input $K$, as mentioned in the objective.
\begin{equation}
\sum_{i} b_i = K, \sum_{i, j, k, l} c_{ij}^{kl} \leq K-1
\end{equation}

\subheadi{Constraint 3.} This constraint enforces that a pair of columns from two tables should only be selected if these two tables are selected to return.
\begin{equation}
2(c_{ij}^{kl} + c_{ji}^{lk}) \leq b_i +  b_j, \forall i, j, k, l
\end{equation}

\subheadi{Constraint 4.}  For simplicity, we assume that tables are always joined via a single-column key, while we leave joining with compound keys in future work. This constraint requires that if two tables are selected to join, they join via a single column from each table.
\begin{equation}
\label{eqn:single_column}
\sum_{k, l} c_{ij}^{kl} \leq 1
\end{equation}

\subheadi{Constraint 5.} This constraint requires each sub-query is covered by at most one column of a table.
\begin{equation}
\sum_k d_{qik} \leq 1, \forall q, i 
\end{equation}

\subheadi{Constraint 6.} This constraint requires that tables used to cover sub-query must be selected to return. $|Q|$ is the number of sub-queries. 
\begin{equation}
\frac{1}{|Q|} \sum_{q} d_{qik} \leq b_i, \forall i, k
\end{equation}

\subsubsection{Sub-query Coverage}
\label{sec:coverage}
The main purpose of measuring fine-grained table-query relevance is to find different columns (may or may not come from the same table) that can map to different parts of the query, so that all information required to answer the query can be covered.
Although we have $\sum_{q, i, k} r_{qik} d_{qik}$ in the objective function to maximize the overall relevance between different parts of the query and columns, it may be biased to aligning a sub-query (weakly) to many columns to maximize the term rather than aligning a sub-query strongly to one column.


Therefore, to encourage strong mappings, we further introduce the following modifications to the MIP formulation. We add an upper limit on the total number of coverage connections between sub-queries and tables that equals to the number of sub-queries. To prevent some sub-queries from not being covered, we introduce the binary decision variable $d_q$ to denote whether or not sub-query $q$ is covered by at least one column. We then add the sum of $d_q$ terms $\alpha \sum_q d_q$ to the objective function. The $\alpha$ coefficient is a knob that controls which strategy to adopt. Large $\alpha$ increases the importance of covering every sub-query covered, which prefers a single table with strong mappings. A low $\alpha$, on the other hand, prefers many tables with weak mappings. The detailed constraints are included in Appendix \ref{app:mip-coverage}.

\subsubsection{Connectedness}
\label{sec:connect}
We consider a graph where tables are nodes and compatibility relationships between two tables are edges. To ensure selected tables are compatible with each other, as discussed at the beginning of Section \ref{sec:mip}, we formulate the following connectedness problem on this graph: the $K$ selected nodes are connected (i.e., any two nodes can be reached by some paths). This problem can be solved by augmenting the original graph and reducing it to a maximum flow problem where connectedness is ensured if and only if the max total flow between source and sink is $K$ \cite{even1975network}.

To ensure connectedness among the selected tables $N$ = $\{T_i | T_i \in C, b_i = 1\}$, we create an augmented graph $G'=(N', E')$ based on $G=(N, E)$, with the following modifications: (1) $N' = N \cup \{source, sink\}$ (2) $E' = E \cup \{ (source, x) \} \cup \{ (n, sink), \forall n \in N \}$ where $x$ is a node in $N$. Then, we define the capacity $h(m, n)$ for each edge between nodes $m$ and $n$ in $N'$. All edges have 0 capacity except the following:
\begin{align*}
h(source, x) = K, h(n, sink) = 1, \forall n \in N\\
h(m, n) = K, \forall m, n \in N
\end{align*}
We include the complete set of constraints in Appendix \ref{app:mip-connect}.




\subsection{Table-Table Relationship Inference} 
\label{sec:cr}
The compatibility between two tables can be represented as the most \emph{joinable} pair of columns between the two tables. To compute the joinablity between two columns, we consider two aspects: column relevance and the likelihood of satisfying the key foreign-key constraint.

A column consists of its schema (header) and instances (rows). Naturally, to compute column relevance, we consider the similarity between its schema and instances. To measure the similarity between column instances, we compute the exact-match overlap (Jaccard similarity).
Given two columns, $c_k$ of $T_i$ and $c_l$ of $T_j$, it is defined as $ \frac{| I(c_k) \cap I(c_l) |}{|I(c_k) \cup I(c_l)|} $
where $I(c_k)$ represents the instances of column $c_k$.
To have a holistic understanding of the column schema, we consider the semantic similarity between two column headers as well as their contextual information, i.e., the corresponding table names and the other columns in the table.
Specifically, we encode each segment using a pre-trained embedding model~\cite{contriever}, and compute the cosine similarity as the similarity score of each segment. Schema similarity is the weighted sum of similarities of all segments. Column relevance is the sum of instance similarity and schema similarity.

We want to further ensure the correctness of joining two columns by considering key foreign-key constraints. For instance, consider table $T_1$ with columns: \{\texttt{student\_id}, \texttt{city}\},  with rows: \{$\langle$1, BOS$\rangle$, $\langle$2, BOS$\rangle$\} and table $T_2$ with three columns: \{\texttt{teacher\_id}, \texttt{student\_id}, \texttt{city}\} with three rows: \{(1, 1, BOS), (2, 1, BOS), (2, 2, BOS)\}. Teachers want to see the cities of students their students are from, so they join $T_1$ and $T_2$, for example, over \texttt{city}. However, this join is incorrect because it creates incorrect teacher-student pairs (e.g., teacher 1 and student 2). A legal join, instead, should correctly connect information from two tables. This occurs when the column $c_k$ with duplicate values (the foreign key) from $T_i$ joins another column $c_l$ (the primary key) from $T_j$ where each of $c_k$'s value is unique.
In short, we want to encourage joins between two columns where at least one of them is a primary key.

To address this, we compute the uniqueness $u$ of both columns $c_k$ and $c_l$ and use $\max \{u(c_k), u(c_l)\}$ to approximate the likelihood of these two columns satisfying the key foreign-key constraint. Uniqueness is defined as the ratio between the number of unique values in a column and the number of instances in the table. If a column is a primary key, its uniqueness score equals to 1, whereas if a column is a foreign key, the uniqueness score should generally be smaller than 1 because they are likely to be repeated (e.g., \texttt{city} column).
Note we take the max of the two uniqueness score because we only need at least one of the two columns to be a primary key.


To summarize, joinablity score $\omega_{ij}^{kl}$ comprises two components: (1) column relevance that includes both instance similarity $j$ and schema similarity $e$ (2) likelihood of satisfying key-foreign key constraint, which is the maximum of the uniqueness $u$ of two columns. Specifically, $\omega_{ij}^{kl} = (e(c_k, c_l) + j(c_k, c_l)) * \max \{u(c_k), u(c_l)\}$.

\section{Evaluation}

In this section, we evaluate the baselines and our re-ranking method on datasets whose questions require multiple tables to answer. The goal is to answer the following two research questions:
(1) To what extent can our re-ranking method, which jointly considers query-table relevance and table-table relevance, improve the existing table retrieval solutions that consider only standard query-table relevance? (2) To what extent can a better retrieval performance due to our re-ranking method enhance the \textit{end-to-end} performance of question answering over tables?


\begin{table*}
\centering

\begin{adjustbox}{max width=\linewidth}
\begin{tabular}{ccccccc|cccccc|ccccccc}
& \multicolumn{6}{c}{\textbf{Top-2}} & \multicolumn{6}{c}{\textbf{Top-5}} & \multicolumn{6}{c}{\textbf{Top-10}} \\
\cmidrule(lr){2-7} \cmidrule(lr){8-13} \cmidrule(lr){14-19}
& \multicolumn{3}{c}{Spider} & \multicolumn{3}{c}{Bird} & \multicolumn{3}{c}{Spider} & \multicolumn{3}{c}{Bird} & \multicolumn{3}{c}{Spider} & \multicolumn{3}{c}{Bird} \\
\cmidrule(lr){2-4} \cmidrule(lr){5-7} \cmidrule(lr){8-10} \cmidrule(lr){11-13} \cmidrule(lr){14-16} \cmidrule(lr){17-19}
& P & R & F1 & P & R & F1 & P & R & F1 & P & R & F1 & P & R & F1 & P & R & F1 \\
\midrule
\multicolumn{19}{l}{\textit{Baselines --- Standard Retrieval Approach}}  \\
\midrule
DTR & 81.4 & 77.4 & 78.9 & 64.9 & 58.9 & 61.3 & 41.0 & 95.8 & 57.1 & 37.2 & 82.9 & 50.9 & 21.3 & 99.3 & 34.9 & 21.5 & 95.4 & 34.8 \\
Contriever & 76.2 & 72.7 & 74.0 & 65.0 & 59.4 & 61.6 & 39.8 & 93.3 & 55.5 & 37.0 & 82.9 & 50.8 & 20.6 & 96.3 & 33.8 & 21.1 & 93.9 & 34.3 \\
\midrule

\multicolumn{19}{l}{\textit{Our methods --- Full Re-ranking
}} \\
\midrule
JAR-F (DTR) & \textbf{87.1} & \textbf{82.9} & \textbf{84.5} & \textbf{76.1} & \textbf{69.3} & \textbf{72.0} & \textbf{41.9} & \textbf{97.8} & \textbf{58.3} & 40.2 & 89.7 & 55.1 & \textbf{21.3} & \textbf{99.7} & \textbf{35.0} & \textbf{21.7} & \textbf{96.6} & \textbf{35.3}\\
JAR-F (Con.) & 82.8 & 79.1 & 80.5 & 73.4 &  67.0& 69.5 & 39.9 & 93.7 & 55.7 & \textbf{40.3} & \textbf{89.9} & \textbf{55.2} & 20.7 & 96.6 & 33.9 &21.5 & 95.4& 34.9 \\

\midrule
\midrule
\multicolumn{19}{l}{\textit{Full Re-ranking with \textbf{gold} key foreign-key constraints}} \\
\midrule
JAR-G (DTR) & 92.2 & 87.9 & 89.6 & 78.0 & 71.1 & 73.8 & 42.4 & 99.2 & 59.1 & 40.9 & 91.3 & 56.0 & 21.4 & 99.9 & 35.1 & 22.1 & 97.8 & 35.8 \\
JAR-G (Con.) & 89.2 & 85.8 & 87.1 & 76.7 & 70.2 & 72.7 & 40.3 & 95.0 & 56.3 & 40.4 & 90.1 & 55.3 & 20.6 & 96.6& 33.8 & 21.8 & 96.5 & 35.3 \\

\midrule
\midrule

\multicolumn{19}{l}{
\textit{Ablation study: Full Re-ranking \textbf{without} table-table relevance}
}\\
\midrule
JAR-D (DTR) & 86.4 & 82.3 & 83.9 & 74.7 & 68.1 & 70.7 & 41.4 & 96.7 &57.6 & 39.5 & 88.4& 54.2 & 21.3 &99.5 & 35.0 & 21.6& 96.2& 35.1\\
JAR-D (Con.) & 81.7 &78.1 & 79.5 & 72.7 & 66.3 & 68.8 & 39.8& 93.3& 55.5 & 39.2 & 87.5 & 53.7 & 20.6 & 96.4 & 33.8 & 21.3 &94.5 &34.5\\

\bottomrule
\end{tabular}
\end{adjustbox}

\caption{Our methods outperform baselines in retrieval performances across all datasets. The top half of the table shows that our full re-ranking attains higher retrieval performances than the standard retrieval methods: DTR and Contriever. Boldface numbers indicate the best performances in the top half of the table. The bottom half of the table shows our approach can effectively use constraints if provided to achieve even better performances. It also shows an ablation suggesting that table-table relevance is essential to the retrieval.
}
\label{tab:retrieval}
\end{table*}

\subsection{Experimental Settings}
\label{sec:exp-setup}

\paragraph{Datasets.}
To the best of our knowledge, there are no existing large-scale open-domain question-answering datasets that consider multiple tables.\footnote{KaggleDBQA~\cite{lee-2021-kaggle-dbqa} has only 26 queries involving multiple tables, which is not a representative dataset for our task.} Therefore, we use text-to-SQL datasets which naturally consider multiple tables for our evaluation, but under an open-domain question-answering setup. In this case, it is required to first retrieve tables from a table corpus, then answer the query by generating an SQL expression. 
Specifically, we run evaluation on two popular text-to-SQL datasets: Spider \cite{yu2018spider} and Bird \cite{li2023llm}.
In these two datasets, tables are organized by topic, and each topic has its own database (around 5.4 tables per database) and queries that can be answered. To make these datasets suitable for our setting, we aggregate tables from all databases to construct a centralized table corpus for each dataset. 
The original datasets contain key foreign-key constraints. However, as described in Section~\ref{sec:intro}, such constraints are not always specified and may need to be inferred during test time. Therefore, we perform evaluations on two setups: with and without the gold key foreign-key constraints provided by the original datasets (if gold constraints are provided, we set the compatibility score of the corresponding pairs of columns to 1).
Moreover, since we focus on queries that require multiple tables to answer, we omit those that only use one table to construct the gold SQL expression. We also remove queries with incorrect answers through human inspection. We provide examples to explain the removal process in Appendix \ref{app:remove}.
The above procedure leads to 443 queries and 81 tables for Spider, and 1095 queries and 77 tables for Bird.

\paragraph{Baselines.}
We set up two baselines for our evaluation. First of all, we consider a strong text retrieval model without explicitly considering the table structure~\cite{contriever}. It is inspired by the study from~\citet{table-text} that table-specific model design may not be necessary for table retrieval. Specifically, we use  Contriever-msmarco\footnote{https://huggingface.co/facebook/contriever-msmarco} to compute the embeddings for the given query and the flattened tables. The ranking of tables is based on the cosine similarities between the query and the tables.
Then, we consider a retrieval method, which is designed for table retrieval, DTR~\cite{opentableqa}.  
Specifically, we finetuned TAPAS-large\footnote{https://huggingface.co/google/tapas-large} following DTR on the tables in the training set (with a 8:2 train-valid split) of each dataset respectively.
Then, the fine-tuned model outputs a score for each question-table pair, which is used for ranking candidate tables.


\paragraph{Environment.}
We use Python-MIP\footnote{https://www.python-mip.com/} package and Gurobi as our MIP solver.
Fine-tuning of DTR models as well as computing coarse- and fine-grained relevance scores are performed on one Tesla V100 GPU. 

\paragraph{Tasks and Metrics.} We conduct the following two types of evaluation.

\subheadi{Retrieval Only Evaluation.}
This evaluation is to directly evaluate to what extent our re-ranking mechanism can improve the baselines with regard to retrieving tables for given queries. We report precision, recall and F1 varying $k$ for our methods and the baselines.

\subheadi{End-to-end evaluation.}
The end-to-end evaluation is to study whether the improved table retrieval can actually help the downstream task, e.g., question answering. Specifically, the ranked tables from various retrieval methods are used to construct prompts (\ref{app:prompt-sql}\footnote{We used one ICL example to teach the model how to generate the output.
}), which will be fed to an LLM to generate an SQL expression to answer the query. In our experiments, we use GPT-3.5 Turbo 1106 (temperature = 0) as our backbone LLM.
To evaluate the end-to-end performance, generated SQL expressions are executed, and we compare the results with those from gold SQL expressions to report the execution accuracy.
Moreover, we report the precision, recall and F1 scores of the tables used in the final SQL expression to demonstrate that better retrieval leads to a better selection of tables by the LLM. 

We use the following notations:
we use \texttt{X} (e.g., DTR) to denote the original baseline approach, \texttt{JAR-D (X)} to denote our re-ranking approach based on \texttt{X} but only considering the fine-grained query-table relevance without table-table relevance, \texttt{JAR-F (X)} to denote our full re-ranking mechanism based on \texttt{X}, and \texttt{JAR-G (X)} to denote the same approach as \texttt{JAR-F (X)} but using the provided gold key foreign-key constraints.

In the following sections, we discuss the results for table retrieval in Section \ref{sec:exp-retrieval} and the end-to-end question answering task in Section \ref{sec:exp-ete}. We also include an ablation study as well as a discussion of the results when gold key foreign-key constraints are available and results under different datasets and numbers of tables in Section \ref{sec:exp-ablation}. 

\subsection{Table Retrieval Performances}
\label{sec:exp-retrieval}
Results of table retrieval are reported in the upper half of Table~\ref{tab:retrieval}, with typical choices of $k$, including 2, 5, and 10. Across different values of $k$, our re-ranking mechanism (JAR-F) can achieve higher retrieval performances compared to baseline retrieval approaches (DTR and Contriever). Specifically, for the Spider dataset, JAR-F (DTR) outperforms DTR by at most 5.6\% in F1 while JAR-F (Contriever) outperforms Contriever by at most 6.5\% in F1. For the Bird dataset, JAR-F (DTR) outperforms DTR by up to 10.7\%, in F1 while JAR-F (Contriever) outperforms Contriever by at most 7.9\%.
The extent of performance improvements decreases as $k$ increases because as the baseline performances increase, it is more difficult to further improve upon them.



\begin{table*}
\centering
\begin{adjustbox}{max width=\linewidth}
\begin{tabular}{cccccc|ccccc}
& \multicolumn{3}{c}{DTR} & JAR-F (DTR) & JAR-G (DTR) & \multicolumn{3}{c}{Contriever} & JAR-F (Con.) & JAR-G (Con.)\\
\cmidrule(lr){2-4} \cmidrule(lr){5-5} \cmidrule(lr){6-6} \cmidrule(lr){7-9} \cmidrule(lr){10-10} \cmidrule(lr){11-11}
& Top-5 & Top-10 & Top-20 & Top-5 & Top-5 & Top-5 & Top-10 & Top-20 & Top-5 & Top-5\\
\midrule
Spider & 46.3 & 47.4 & 48.1 & \textbf{50.6} & 52.2 & 43.5 & 43.1 & 44.7 & 47.4 & 48.3 \\
Bird & 30.4 & 34.8 & 31.5 & \textbf{35.8} & 36.9 & 29.7 & 30.6 & 32.9 & 35.1 & 36.2\\
\bottomrule
\end{tabular}
\end{adjustbox}

\caption{Our re-ranking mechanism achieves considerable improvements compared to baseline approaches on both Spider and Bird datasets in terms of end-to-end execution accuracy.}
\label{tab:ete-ex}
\end{table*}

\begin{table}
\centering

\begin{adjustbox}{max width=\linewidth}
\small
\begin{tabular}{ccccccc}
& \multicolumn{6}{c}{\textbf{Top-5}}  \\
\cmidrule(lr){2-7}
& \multicolumn{3}{c}{Spider} & \multicolumn{3}{c}{Bird} \\
\cmidrule(lr){2-4} \cmidrule(lr){5-7}
 & P & R & F1 & P & R & F1  \\
\midrule
\multicolumn{6}{l}{\textit{Full Re-ranking}} \\
\midrule
JAR-F (DTR) & \textbf{77.2} &\textbf{77.4} &\textbf{77.0} & \textbf{65.6} & \textbf{66.1} & \textbf{65.3} \\
JAR-F (Con.) & 75.4 & 74.8 & 74.7 & 64.7 & 65.4 & 64.7 \\
\midrule
\multicolumn{6}{l}{\textit{Baselines}}  \\
\midrule
& \multicolumn{6}{c}{\textbf{Top-5}}  \\
\cmidrule(lr){2-7}
DTR & 75.8  & 75.6 & 75.2 & 59.5 & 59.2 & 58.8  \\
Contriever & 71.9 & 71.4 & 71.3 & 55.9 & 55.5 & 55.2  \\
 & \multicolumn{6}{c}{\textbf{Top-10}}\\
 \cmidrule(lr){2-7}
DTR & 77.6 & 76.7 & 76.7 & 65.4 & 65.2 & 64.7\\
Contriever & 74.1 & 73.2 & 73.2 & 61.8 & 62.5 & 61.7\\
& \multicolumn{6}{c}{\textbf{Top-20}} \\
\cmidrule(lr){2-7}
DTR & 76.0 & 74.8 & 75.1 & 64.5 & 64.8 & 64.2  \\
Contriever & 74.4 & 73.6 & 73.6 & 64.7 & 64.4 & 64.1\\

\bottomrule
\end{tabular}
\end{adjustbox}


\caption{Our approach attains significant gains compared to baseline approaches on both Spider and Bird datasets in end-to-end retrieval performances.}
\label{tab:ete-retrieval}
\end{table}

\subsection{End-to-end Performances}
\label{sec:exp-ete}



In this section, we want to show that a better retrieval performance can translate to a better end-to-end performance on the downstream task, e.g., question answering through text-to-SQL. In table~\ref{tab:ete-ex}, we report the execution accuracy of SQL expressions, and in table~\ref{tab:ete-retrieval}, we report the ``retrieval performance'' of tables by the LLM when generating the final SQL expressions.

In table~\ref{tab:ete-ex}, we observe that when feeding the same number of tables (top-5) to the LLM, our full re-ranking solutions can significantly outperform the baselines. Since our re-ranking mechanism is an additional step between the standard table retrieval and LLM response generation, we also want to investigate whether the LLM can do some level of re-ranking. Therefore, we also compare against the baseline which feeds 20 tables (the same number of tables we feed to our re-ranking method) to the LLM.
Table \ref{tab:ete-ex} shows, on average across both models, an 2.6\% and 3.3\% improvement on execution accuracy for Spider and Bird respectively.
We then further investigate feeding LLM with top-10 tables, which represents a balance between precision and recall compared to top-5 and 20, and we can observe similar improvements. All these results demonstrate the effectiveness of our re-ranking mechanism. 




In table~\ref{tab:ete-retrieval}, we further investigate the table retrieval performance by the LLM with different input of tables produced either by the baselines or our methods.
As shown in the table,  our approach with full re-ranking outperforms baselines across all datasets and numbers of input tables for baseline approaches. Compared to the best overall baseline (top-20), our approach still, on average, outperforms it by 1.5\% and 0.85\% in F1 for Spider and Bird respectively. It demonstrates that our re-ranking mechanism actually improves the end-to-end performance by helping the LLM select the correct tables.


\subsection{Discussion}
\label{sec:exp-ablation}


\paragraph{Re-ranking without table-table relevance.} As shown in Table \ref{tab:retrieval}, performing re-ranking using fine-grained query-table relevance (JAR-D) already outperforms the baseline approaches (DTR and Contriever), demonstrating the effectiveness of fine-grained query-table relevance. Further considering table-table relevance can further enhance retrieval performances as JAR-F generally outperforms JAR-D. This suggests that join compatibility is helpful. We conclude that both fine-grained and table-table relevance are essential to better retrieval performances.

\paragraph{Full re-ranking with gold key foreign-key constraints.} Table \ref{tab:retrieval} and Table \ref{tab:ete-ex} suggest that complementing full re-ranking with the gold key foreign-key constraints (JAR-G) leads to better performances compared to full re-ranking (JAR-F). This is because gold join relationships can generate higher-quality compatibility scores and our re-ranking mechanism can effectively leverage better scores to generate better outputs.

\begin{table}
\centering

\begin{adjustbox}{max width=\linewidth}
\begin{tabular}{ccccccc|cccccc}
& \multicolumn{6}{c}{Queries with 2 tables} &  \multicolumn{6}{c}{Queries with $\geq 3$ tables} \\
\cmidrule(lr){2-7} \cmidrule(lr){8-13}
& \multicolumn{3}{c}{Spider} & \multicolumn{3}{c}{Bird} & \multicolumn{3}{c}{Spider} & \multicolumn{3}{c}{Bird} \\
\cmidrule(lr){2-4} \cmidrule(lr){5-7} \cmidrule(lr){8-10} \cmidrule(lr){11-13} 
 & P & R & F1 & P & R & F1 & P & R & F1 & P & R & F1  \\
\midrule
\multicolumn{13}{l}{Table retrieval performances (Top-5)}\\
\midrule
\multicolumn{6}{l}{\textit{DTR}} \\
\midrule
JAR-D & 38.8 & 97 & 55.4 & 36.1	&90.3	&51.6 & 58.9&	94.7&	72.5 & 50.8	&82.7	&62.8 \\
JAR-F & 39.2&	98&	56& 36.5&	91.3	&52.2 & 59.6&	95.9	&73.4 & 52	&84.6&	64.3 \\
Diff. & 0.4 & 1.0 & 0.6 & 0.4 & 1.0 & 0.6 & \textbf{0.7} & \textbf{1.2} & \textbf{0.9} & \textbf{1.2} & \textbf{1.9} & \textbf{1.5}\\
\midrule
\multicolumn{6}{l}{\textit{Contriever}} \\
\midrule
JAR-D & 37.9&	94.8&	54.2 &35.9	&89.7	&51.2 & 52.3	&83.6&	64.2 & 49.8	&80.7	&61.5 \\
JAR-F & 38	&94.9&	54.2 & 36.6&	91.6	&52.3 & 53	&85.1&	65.2 & 52&	84.5	&64.3 \\
Diff. & 0.1 & 0.1 & 0.0. & 0.7 & 1.9 & 1.1 & \textbf{0.7} & \textbf{1.5} & \textbf{1.0} & \textbf{2.2} & \textbf{3.8} & \textbf{2.8} \\

\bottomrule
\end{tabular}
\end{adjustbox}


\caption{Our approach is the most effective on more challenging and realistic queries involving more tables.}
\label{tab:num_table}
\end{table}

\paragraph{Retrieval performances under different datasets and numbers of tables.} Table \ref{tab:num_table} shows that table-table relevance contributes (as measured by the difference between the retrieval performances of JAR-F and JAR-D) more to retrieval performances for (1) queries involving more tables and (2) Bird compared to Spider. 
Since our focus is to improve table-related downstream tasks involving multiple tables, the fact that table-table relevance provides the most benefits with a higher number of tables should be expected. Moreover, we find that queries in the Bird dataset are more challenging and realistic because they involve more bridging tables (e.g., the \textit{Disp} table as mentioned in Figure \ref{fig:intro_exp}) that are less directly relevant to the user question. They also contain more confusing tables (tables with similar schema, such as \textit{Client} (\texttt{client\_id}) and \textit{Customers} (\texttt{CustomerID})). These features resemble what exists in real life and table-table relevance can address these more challenging issues, demonstrating its effectiveness.



\section{Related Work}
One line of related work is text-to-SQL~\cite{Zhong2017Seq2SQLGS,yu2018spider}. Thanks to recent advances in LLM, SQL queries (including tables) corresponding to a natural language question or query can be generated through prompting engineering~\cite{nan2023enhancing,liu2023comprehensive} or fine-tuning~\cite{10.14778/3641204.3641221, wang2024dbcopilot} with LLMs. Although, in this paper, we evaluate our method on the aggregated version of text-to-SQL datasets, we target a general multi-table retrieval problem that requires the consideration of multiple (joinable) tables. In other words, we do not limit our work to specific downstream tasks (e.g., text-to-SQL for a given database).

To the best of our knowledge, we are the first work to propose the problem of join-aware multi-table retrieval for complex questions, and our paper encourages considering table relationships for retrieval during inference time. One related direction from the literature is fusion retrieval~\cite{chen2020open}. The basic idea of fusion retrieval is based on an ``early fusion'' strategy to group relevant heterogeneous data (including text and tables) before retrieval.
The early fusion process aims to fuse a table segment and relevant passages into a group. It is implemented by linking entities mentioned in a table segment to the appropriate passages, similar to traditional document expansion based on individual entities.
Although this method can also capture the relationship between table and text, it cannot capture key foreign-key relationships among multiple tables. Therefore, when answering aggregation questions, false-positive tables are very likely to be introduced due to the expansion through individual entities. 
Furthermore, tables do not have to be connected by columns of entities, and the foreign key column can be just as simple as an id column.
Last but not least, real questions may require more than 2 tables to answer, and therefore bridging tables may be needed to connect tables. Sometimes, the bridging tables can also cross multiple hops. In this case, tables introduced by an "early fusion" strategy will grow exponentially, which may not help the downstream LLM to answer the question correctly. In this work, we also aim to narrow down the number of correlated tables in a limited ranked list, so that they can best help the LLM to answer the question correctly.

\section{Conclusion}
Table retrieval is critical for open-domain question answering, fact-checking, and, more generally, retrieval-augmented generation that relies on table information. However, existing work primarily focuses on single-table retrieval without considering real-life situations where tables are typically normalized. Thus, multiple tables need to be joined to produce sufficient information. 
Works that consider multi-table retrieval only consider query-table relevance. 
In this paper, we propose a re-ranking method based on mixed-integer programming to select the best set of tables based on joint decisions about relevance and compatibility. 
Our experiments demonstrated that our approach outperforms baselines in standard retrieval performance and end-to-end evaluation on the question-answering downstream task.

\section{Limitations}
The primary purpose of this paper is to remind the community that table retrieval is still not a solved problem. For complex queries, retrieving multiple (joinable) tables may be required. The proposed MIP-based reranking mechanism in this paper takes a first step to consider join relationships among tables during the inference time. Though MIP may have scalability issues and can be sensitive to the problem data, it provides a flexible framework to model constraints and objectives.


Furthermore, key-foreign-key relationships may not be the only connection among multiple tables for questions in real-life scenarios. 
Some questions may ask about values of the same column across multiple tables.

We believe exploring a more scalable solution, as well as considering different types of connections among tables in the corpus, would be interesting future work.

\section*{Acknowledgments}

We thank Mike Cafarella and \c{C}a\u{g}atay Demiralp for their constructive feedback and discussion on this project. We thank the MIT SuperCloud and Lincoln Laboratory Supercomputing Center for providing computing resources that have contributed to the research results reported in this paper. We gratefully acknowledge the support of the DARPA ASKEM project (Award No. HR00112220042), ONR Contract N00014-23-1-2364, and the Croucher Scholarship.

\bibliography{anthology,custom}

\appendix

\section{Prompts}
\label{app:prompt}

\subsection{Decomposition}
\label{app:prompt-decomp}
We use the following prompt to ask GPT-3.5 Turbo to decompose a question into (concept, attribute) pairs

\texttt{I'm going to ask you a question. I want you to decompose it into a series of non-composite noun concepts and attributes. If you are uncertain about concepts, only output attributes. You should wrap each concept and attribute in <sub\_c></sub\_c> tags. Once you have all the concepts you need to cover the question, output <FIN></FIN> tags.
\\
Let's go through some examples together.\\
Question: For movies with the keyword of "civil war", calculate the average revenue generated by these movies.\\
\\
Answer:\\
<sub\_c>movies:keyword</sub\_c>\\
<sub\_c>movies:revenue</sub\_c>\\
<FIN></FIN>\\
\\
Question: How many customers have a credit limit of not more than 100,000 and which customer made the highest total payment amount for the year 2004?\\
\\
Answer:\\
<sub\_c>customers:credit limit</sub\_c>\\
<sub\_c>customers:payment amount</sub\_c>\\
<sub\_c>year</sub\_c>\\
<FIN></FIN>\\
\\
Question: What is the aircraft name for the flight with number 99?\\
\\
Answer:\\
<sub\_c>aircraft:name</sub\_c>\\
<sub\_c>flight:number</sub\_c>\\
<FIN></FIN>\\
\\
Question: On which day has it neither been foggy nor rained in the zip code of 94107?\\
\\
Answer:\\
<sub\_c>zip code</sub\_c>\\
<sub\_c>weather</sub\_c>\\
<FIN></FIN>\\
\\
Question: What is the id of the trip that started from the station with the highest dock count?\\
\\
Answer:\\
<sub\_c>trip:id</sub\_c>\\
<sub\_c>station:dock count</sub\_c>\\
<FIN></FIN>\\
\\
Question: ...\\
\\
Answer:}

\subsection{SQL generation}
\label{app:prompt-sql}
We use the following prompt to ask GPT 3.5 Turbo 1106 to generate SQLs:

\noindent \texttt{// TASK INSTRUCTION\\
Generate SQL given the question, tables, and external knowledge to answer the question correctly. First, identify tables with relevant columns. Then, join these tables using only columns in the tables. Finally, decide which columns to return in the SQL to answer the original question. When returning columns, please specify the tables associated with it to prevent ambiguity. Think step by step.\\
\\
// 1-SHOT PSEUDO EXAMPLE\\
CREATE TABLE singer(\\
\indent singer\_id TEXT: <instance 1>, ...,\\
\indent  nation TEXT: <instance 1>, ...,\\
\indent  sname TEXT: <instance 1>, ...,\\
\indent  dname TEXT: <instance 1>, ...,\\
\indent  cname TEXT: <instance 1>, ...,\\
\indent  age INTEGER: <instance 1>, ...,\\
\indent  year INTEGER: <instance 1>, ...,\\
\indent  birth\_year INTEGER: <instance 1>, ...,\\
\indent  salary REAL: <instance 1>, ...,\\
\indent  city TEXT: <instance 1>, ...,\\
\indent  phone\_number INTEGER: <instance 1>, ...,\\
\indent  tax REAL: <instance 1>, ...)\\
\\
External Knowledge: age = year - birth\_year\\
Question: How many singers in USA who is older than 27?\\
Answer: SELECT COUNT(*) FROM singer WHERE year - birth\_year > 27;\\
\\
// NEW INPUT \\
CREATE TABLE <table1>(\\
\indent  <Column1> <Type1>: <instance 1>, ...,\\
\indent  <Column2> <Type2>: <instance 1>, ...)\\
...\\
\\
\\
External knowledge: ... \\
Question: ...\\
Answer:
}

\section{MIP formulation}
\subsection{Coverage quality}
\label{app:mip-coverage}
\begin{equation}
\alpha \sum_q d_q
\end{equation}

\begin{equation}
d_q \in \{0, 1\} 
\end{equation}
\texttt{(integral constraints)}

\begin{equation}
d_q \leq \sum_{i, k} d_{qik}, \forall q
\end{equation}
\texttt{(A sub-query is covered if it is covered by at least one column)}

\begin{equation}
\sum_{q,i,k} d_{qik} \leq |Q|
\end{equation}
\texttt{(total number of question-table connections do not exceed the total number of sub-queries)}

\subsection{Connectedness}
\label{app:mip-connect}


We introduce these additional variables: $s_i$ denotes the amount of flow from source to table $i$; $t_i$ denotes the amount of flow from table $i$ to sink; $e_i$ denotes whether or not to give an initial flow of $k$ to table $i$; $f_{ij}^{kl}$ denotes amount of flow from column $k$ of table $i$ to column $l$ of table $j$.

\begin{equation}
e_i \in \{0, 1\}, 0 \leq t_i \leq 1, s_i \geq 0, f_{ij}^{kl} \geq 0
\end{equation}
 \texttt{(variables constraints)}

\begin{equation}
s_i + \sum_{j, k, l} f_{ji}^{lk} = \sum_{j, k, l}f_{ij}^{kl} + t_i, \forall i
\end{equation}
\texttt{(flow balance constraint)}

\begin{equation}
\sum_{i} t_i = K
\end{equation}
\texttt{(the amount of flow into the sink should be $K$)}

\begin{equation}
\sum_{i} e_i = 1, e_i \leq M b_i, s_i = K e_i, \forall i
\end{equation}
\texttt{(select one chosen table to give all $K$ units of flow)}

\begin{equation}
\frac{1}{k} f_{ij}^{kl} \leq c_{ij}^{kl} \leq M f_{ij}^{kl}
\end{equation}
\texttt{(linking constraint between flow and column selection)}



\section{Dataset processing}
\label{app:remove}

We remove questions if their corresponding gold SQL expressions involve table joins that generate incorrect instances. This typically happens when a non-unique foreign key of a table is joined with another non-unique foreign key of another table. We illustrate one example from each dataset:

\paragraph{Spider}
Consider the following two tables:

\textit{AREA\_CODE\_STATE} (\texttt{area\_code}, state)

\textit{VOTES} (\texttt{vote\_id}, phone\_number, state, ...)

To obtain information about the vote and the area code of each vote, these two tables are joined using ``state''. However, a state can have multiple areas and thus its value is not unique in \textit{AREA\_CODE\_STATE}. Therefore, joining over the state column creates incorrect instances where a vote in, for example, Area 1 of a state is connected to Area 2 of the same state simply because they are in the same state.

\paragraph{Bird}

Consider the following two tables:

\textit{Account} (\texttt{account\_id}, district\_id, frequency, date)

\textit{Client} (\texttt{client\_id}, gender, birth date, district\_id)

To obtain information about client and their accounts, these two tables are joined using \texttt{district\_id}. However, this join creates incorrect instances, similar to the case above in Spider. For example, if two clients live in the same district and opened their accounts in the same district, such a join creates instances involving each client with the other client's account which is incorrect.

\end{document}